\documentclass[RNAAS]{aastex62}

\begin{document}

\title{Discovery of 13 New Orbital Periods for Classical Novae}

%% Note that the corresponding author command and emails has to come
%% before everything else. Also place all the emails in the \email
%% command instead of using multiple \email calls.
\correspondingauthor{Bradley E. Schaefer}
\email{schaefer@lsu.edu}

\author{Bradley E. Schaefer}
\altaffiliation{Department of Physics and Astronomy, Louisiana State University, Baton Rouge Louisiana, 70803 USA}

%% Note that RNAAS manuscripts DO NOT have abstracts.
\begin{abstract}

I report new orbital periods ($P$) for 13 classical novae, based on light curves from {\it TESS}, {\it AAVSO}, and other public archives.  These new nova periods now constitute nearly one-seventh of all known nova periods.  Five of my systems have $P$$>$1 day, which doubles the number of such systems that must have evolved companion stars.  (This is simply because ground-based time series have neither the coverage nor the stability required to discover these small-amplitude long periods.)  V1016 Sgr has a rare $P$ below the period gap, and suddenly becomes useful for current debates on evolution of novae.  Five of the novae (FM Cir, V399 Del, V407 Lup, YZ Ret, and V549 Vel) have the orbital modulations in the tail of the eruption after the transition phase.  Soon after the transition, YZ Ret shows a unique set of aperiodic diminishing oscillations, plus YZ Ret shows two highly-significant transient periods, 1.1\% and 4.5\% longer than the orbital period, much like for the superhump phenomenon.  I also report an optical 591.27465 second periodicity for V407 Lup, which is coherent and must be tied to the white dwarf spin period.  The new orbital periods in days are 0.1883907$\pm$0.0000048 for V1405 Cas, 3.4898$\pm$0.0072 for FM Cir, 0.162941$\pm$0.000060 for V339 Del, 3.513$\pm$0.020 for V407 Lup, 1.32379$\pm$0.00048 for V2109 Oph, 3.21997$\pm$0.00039 for V392 Per, 0.1628714$\pm$0.0000110 for V598 Pup, 0.1324539$\pm$0.0000098 for YZ Ret, 0.07579635$\pm$0.00000017 for V1016 Sgr, 7.101$\pm$0.016 for V5583 Sgr, 0.61075$\pm$0.00071 for V1534 Sco, 0.40319$\pm$0.00005 for V549 Vel, and 0.146501$\pm$0.000058 for NQ Vul.

\end{abstract}

%% Start the main body of the article. If no sections in the 
%% research note leave the \section call blank to make the title.
\section{Orbital Periods of Classical Novae} 

The single most important number for any classical nova is its orbital period, $P$.  With this, the nova community has a long-running and vast program seeking to discover $P$, where now the 337 confirmed classical novae known to date in our galaxy have produced orbital periods for roughly 94 systems.

With the launch of the {\it Kepler} and {\it TESS} spacecrafts, new opportunities arise to discover new nova periods.  The basis is that all nova binaries (excepting only those viewed nearly face-on) will show photometric modulation on the orbital period.  Both spacecrafts have incredibly good photometric accuracy and stability, plus awesome coverage with short exposure times (variously 20 seconds to 30 minutes) and {\it continuous} `movies' for many days (variously 27 to 90 days).  With coverage over much of the sky, most old novae (bright enough in quiescence  for a useable signal) should display a photometric periodicity that will almost surely be the orbital period.  So my plan is to use public domain light curves from {\it TESS} and {\it Kepler}, plus a wide variety of other public datasets, to seek all available orbital periods.

\section{New Orbital Periods} 

I have examined 22 old novae for which no confident $P$ is known and for which I could extract two useable data sets.  The primary source was from the {\it TESS}, due to its depth and nearly-full-sky coverage.  The {\it K2} mission for the {\it Kepler} spacecraft provided good coverage of the galactic center region.  The other workhorse dataset is the light curves collected by the {\it American Association of Variable Star Observers} ({\it AAVSO}).  I also made use of light curves from {\it All Sky Automated Survey} ({\it ASAS}) and {\it ASAS for Supernovae} ({\it ASAS-SN}).  The general procedure was to clean the data of artifacts, detrend as needed, then run a discrete Fourier transform to recognize coherent periodicities.  For each significant period, I fit a light curve template (usually a sinewave) in a chi-square minimization, with the error bars determined from the minimum chi-square plus unity.  Formal significances were evaluated from both the Fourier transform and the chi-square.

A primary issue is to establish high confidence in the period, as the literature has many dubious, contradictory, and insignificant period claims.  (1) The periodicity must be significant, even when accounting for the number of independent trial periods in the Fourier transform.  (2) My primary check is that I require the same periodicity to appear significantly in two independent data sets.  This strongly guards against many instrumental artifacts, as well as random alignments of flickers or dips.  For two novae, I find significant and good-looking periodicities in only one data set, and no other data set exists that can provide an independent test for such a low amplitude modulation, with these being explicitly separated below the line in the table.  That is, these two periods satisfy the usual standards for acceptance, but not my stricter requirement of appearing significantly in {\it two} independent data sets.

I have found 14 new periods, 13 of which are orbital, all tabulated in the table.  The two data sets are given with the start year and the number of magnitudes/fluxes.  All have a phased light curve shaped like a sinewave, pointing to the expected irradiation effects present in all novae light curves, and thus giving the orbital period.  V5583 Sgr also displays shallow primary and secondary eclipses.  The photometric modulations of FM Cir, V399 Del, V407 Lup, YZ Ret, and V549 Vel are seen in the tail of the eruption (after the transition).  Here are comments on individual novae:

{\bf V1405 Cas:}  Before the nova, the counterpart had been identified as CzeV 3217 Cas, with a reported $P$=0.376938 days and a classification as EW.  However, their original data is not adequate (in either coverage or accuracy) to distinguish between an EW binary and an ordinary cataclysmic binary with a near sinewave light curve at half the EW period.  The {\it TESS} pre-eruption light curve does not show a shape for an EW binary, but rather shows a roughly sinusoidal light curve with flickering superposed.  So the system is an ordinary cataclysmic variable with the real orbital period of 0.1883907 days.

{\bf V407 Lup:}  The prior claim for the orbital period of 3.57 hours from the UVOT light curve (Aydi et al. 2018) was never significant, and is now strongly disproven with the TESS light curve.  The prior claim for the optical periodicity of 565 seconds (Aydi et al. 2018) was mistaken.  The observed period of 591.27465$\pm$0.00023 seconds is coherent and must be tied to the white dwarf spin period, perhaps as a sideband.

{\bf V392 Per:}  The claimed periodicity of Munari et al. (2020) was selected from his Fourier transform with a `forest' of many peaks with similar power, produced by only 15 brief visits (each with an RMS scatter roughly equal to half the amplitude) .  The greatly different claimed periodicity of Schmidt (2020b) suffered all the same problems. 

{\bf YZ Ret:}  In the three months following the transition, two highly significant periodicities (around $P$=0.1384 and $P$=0.1339 days) are seen to come and go.  The differences from the tabulated pre-eruption period is greatly too large to represent any change in the orbital period, so the modulation must be related to the accretion disk, perhaps as some precessing out-of-round shape, perhaps like the superhump phenomenon.  For the first 15 days, the light curve also has aperiodic dips separated by 0.3--1.0 days with amplitudes descending from 0.1 mag to under 0.003 mag.  The only precedent for this phenomenon is the aperiodic dips seen just after the transition in the reforming disk of the recurrent nova U Sco (Schaefer et al. 2010).

{\bf V1016 Sgr:}  The one-day alias period of 0.070456175$\pm$0.00000017 days is possible.  From 12 archival sky photographs at Harvard, I have measured an average Johnson B magnitude of 15.5 from {\it before} its 1899 eruption.  With the {\it Gaia} distance, the pre-eruption quiescent absolute magnitude is +2.3 mag.  From many sources, the post-eruption B magnitude is 15.3.  This result contradicts the strong prediction from the Hibernation model of evolution that $P<2$ hour systems must have the post-eruption brightness (after the light curve goes flat) more than 5.5 mag {\it fainter} than immediately before eruption (Hillman et al. 2020).  Further, the Hibernation model requires that the $P<2$ hour novae have absolute magnitude of close to +6.5 mag, in contradiction by 50$\times$ from observations.

{\bf V5583 Sgr:}  The usual irradiation-sinewave has superposed shallow primary and secondary eclipses.  This characteristic shape is followed closely for four cycles, and this is our guarantee that we are not just seeing flickering.

I have searched for orbital periods of V679 Car, V906 Car, V705 Cas, V1405 Cen, V827 Her, DK Lac, V1112 Per, V5666 Sgr, and V5667 Sgr from 0.04--10 days, all with no significant period.  V1405 Cen has five dips producing an apparent period of 2.73 days, but the dips are too few to generate any useable confidence.  The claimed periodicities for V2891 Cyg (Schmidt 2020a) and V1112 Per (Schmidt 2021) are both greatly insignificant as tiny noise peaks amongst a forest of similar noise peaks in the presented periodograms, these two claimed periodicities are emphatically absent from the {\it TESS} light curves taken at the same time to very deep limits, while the photometry is all from times when the shell is optically thick so any photometric modulations from the inner binary are completely hidden.  V5666 Sgr has an apparent period of P=6.62 days, but this looks to be just random flickering.

%As part of this same study, I have found the orbital periods of KT Eri and V2487 Oph, but these are parts of complex situations, so these periods are reserved for separate papers.  With this, 15 out of 24 novae have revealed their $P$, despite most of them having already been intensively examined previously.  This is a testimonial for the power of {\it TESS} and {\it AAVSO}.

\movetabledown=35mm
\begin{rotatetable}
\begin{deluxetable*}{llllllll}
\tablecaption{New Periods for Classical Novae}
\tablehead{
\colhead{Nova} & \colhead{Year} & \colhead{P (days)} & \colhead{Minimum (BJD)} & \colhead{Amp (mag)} & \colhead{Dataset 1 (year, \#)}& \colhead{Dataset 2 (year, \#)}
}
\startdata
\hline
V1405 Cas	&	2021	&	0.1883907	$\pm$	0.0000048	&	2458859.0688	$\pm$	0.0021	&	0.0027	&	TESS (2019, 1831)	&	TESS (2020, 1167)	\\
FM Cir	&	2018	&	3.4898	$\pm$	0.0072	&	2458624.6918	$\pm$	0.0280	&	0.0033	&	TESS (2019, 1073)	&	TESS (2019, 1173)	\\
V339 Del	&	2013	&	0.162941	$\pm$	0.000060	&	2458696.1286	$\pm$	0.0028	&	0.020	&	TESS (2019, 7109)	&	TESS (2019, 7058)	\\
V407 Lup	&	2016	&	0.0068434566	$\pm$	0.0000000026	&	2458152.00076	$\pm$	0.00008	&	0.026	&	AAVSO (2017, 5626)	&	AAVSO (2018-9, 8718)	\\
V2109 Oph	&	1969	&	1.32379	$\pm$	0.00048	&	2457693.0927	$\pm$	0.0077	&	0.199	&	K2 (2016, 1048)	&	K2 (2016, 2027)	\\
V392 Per	&	2018	&	3.21997	$\pm$	0.00039	&	2459135.4580	$\pm$	0.0095	&	0.122	&	AAVSO (2019-21, 28725)	&	TESS (2019, 1150)	\\
V598 Pup	&	2007	&	0.1628714	$\pm$	0.0000110	&	2459228.3167	$\pm$	0.0010	&	0.038	&	TESS (2020-1, 17456)	&	TESS (2021, 16891)	\\
YZ Ret	&	2020	&	0.1324539	$\pm$	0.0000098	&	2458408.0161	$\pm$	0.0013	&	0.031	&	TESS (2018, 1201)	&	TESS (2018, 853)	\\
V1016 Sgr	&	1899	&	0.07579635	$\pm$	0.00000017	&	2453447.0357	$\pm$	0.0020	&	0.138	&	ASAS (2001-9, 446)	&	ASAS-SN (2016-8, 159)	\\
V1534 Sco	&	2014	&	0.61075	$\pm$	0.00071	&	2458640.9536	$\pm$	0.0099	&	0.022	&	TESS (2019, 348)	&	TESS (2019, 303)	\\
V549 Vel	&	2017	&	0.40319	$\pm$	0.00005	&	2458556.1609	$\pm$	0.0009	&	0.052	&	TESS (2019, 588)	&	TESS (2019, 560)	\\
NQ Vul	&	1976	&	0.146501	$\pm$	0.000058	&	2458697.0328	$\pm$	0.0034	&	0.0025	&	TESS (2019, 597)	&	TESS (2019, 537)	\\
\hline
V407 Lup	&	2016	&	3.513	$\pm$	0.020	&	24588641.8788	$\pm$	0.0515	&	0.0029	&	TESS (2019, 1111)	&	...	\\
V5583 Sgr	&	2009	&	7.101	$\pm$	0.016	&	2458671.7680	$\pm$	0.0225	&	0.009	&	TESS (2019, 1216)	&	...	\\
\enddata
\end{deluxetable*}  
\end{rotatetable}

\end{document}